\documentclass[prl,amsmath,amssymb,notitlepage,twocolumn,showpacs]{revtex4-1} 

\usepackage{graphicx}
\usepackage[]{hyperref}
\usepackage{color}

\newcommand{\beq}{\begin{equation}} 
\newcommand{\eeq}{\end{equation}}

\newcommand{\Srf}{$^{87}$Sr}
\newcommand{\Srb}{$^{88}$Sr}

\newsavebox{\Wbox}
\savebox{\Wbox}{\scalebox{0.6}{W}}

\definecolor{correction}{RGB}{255,0,0}

\begin{document}

\title{Test of Einstein Equivalence Principle for 0-spin and half-integer-spin atoms:\\ Search for spin-gravity coupling effects}
 
\author{M.~G.~Tarallo}
\altaffiliation{Present address: Department of Physics, Columbia University, 538 West 120th Street, New York, NY 10027-5255, USA}
\author{T.~Mazzoni}
\author{N.~Poli}
\author{D.~V.~Sutyrin}
\author{X.~Zhang}
\altaffiliation{Also: International Center for Theoretical Physics (ICTP), Trieste, Italy }
\author{G.~M.~Tino} 
\email{Guglielmo.Tino@fi.infn.it}

\affiliation{Dipartimento di Fisica e Astronomia and LENS -- Universit\`a di Firenze, INFN -- Sezione di Firenze,\\ Via Sansone 1, 50019 Sesto Fiorentino, Italy}

\date{\today}

\begin{abstract} 
We report on a conceptually new test of the equivalence principle performed by measuring the acceleration in Earth's gravity field of two isotopes of strontium atoms, namely, the bosonic \Srb~isotope which has no spin vs the fermionic \Srf~isotope which has a half-integer spin. The effect of gravity upon the two atomic species has been probed by means of a precision differential measurement of the Bloch frequency for the two atomic matter waves in a vertical optical lattice. We obtain the values $\eta = (0.2\pm 1.6)\times10^{-7}$ for the E\"otv\"os  parameter and $k=(0.5\pm1.1)\times10^{-7}$ for the coupling between nuclear spin and gravity. This is the first reported experimental test of the equivalence principle for bosonic and fermionic particles and opens a new way to the search for the predicted spin-gravity coupling effects.
\end{abstract}

\keywords{}
\pacs{
03.75.Dg, 
04.80.Cc, 
37.25.+k, 
37.10.Jk 
}

\maketitle

{The} Einstein equivalence principle (EP) is at the heart of general relativity, the present theory of gravity~\cite{Misner}. In its so called \emph{weak form}, corresponding to the universality of free fall, it goes back to Galileo Galilei's idea that the motion of a mass in a gravitational field is independent of its structure and composition. Violations of the EP are expected  in attempts to unify general relativity with the other fundamental interactions and in theoretical models for dark energy in cosmology~\cite{Colladay1997,Damour2002} {as well as in extended theories of gravity~\cite{Capozziello2011}}.

The most stringent experimental limits for the EP to date come from two methods: the study of the motion of moons and planets  and the use of torsion balances~\cite{Will2006}. In recent years, experiments based on atom interferometry~\cite{Cronin2009,Tino2014} compared the fall in the Earth's gravitational field of two Rb isotopes \cite{Fray:2004p65,Bonnin13} and Rb vs K \cite{Schlippert2014} reaching a relative precision of about $10^{-7}$. Tests of EP were carried out in which the measurement of Earth's gravity acceleration with an atom interferometer was compared with the value provided by a classical gravimeter~\cite{Peters1999, Poli2011}. A much higher precision will be achieved in future experiments with atom interferometers that are planned on ground \cite{Dimopoulos2007} and in space~\cite{Tino2013,Ste-quest}. The possibility of tests with atom interferometry for matter vs antimatter was also investigated \cite{Kellerbauer2008,Hohensee2013}. The interest of using atom
 s is indeed not only to improve the limits reached by classical tests with macroscopic bodies, but mostly in the possibility to perform qualitatively new tests with ``test masses'' having well defined properties, e.g. in terms of spin, bosonic or fermionic nature, and proton-to-neutron ratio.

Possible spin-gravity coupling, torsion of space-time, and EP violations have been the subject of extensive theoretical investigation (see, for example, ~\cite{Hehl76,Peres1978,Mashhoon2000,Obukhov2001,Bini2004,Capozziello2001,Ni2010}).
Experimental tests were performed based on macroscopic test masses \cite{Heckel2008,Ni2010}, atomic magnetometers \cite{Venema1992,Kimball2013}, atomic clocks \cite{Wineland1991}. In~\cite{Fray:2004p65}, a differential free fall measurement of atoms in two different hyperfine states was also performed. Possible differences in gravitational interaction for bosonic and fermionic particles were also discussed~\cite{Barrow,Herrmann2012} and efforts towards experimental tests with different atoms are underway~\cite{Varoquaux2009,Herrmann2012}.

\begin{figure}[tb]
\begin{center}
\includegraphics[width=8. cm]{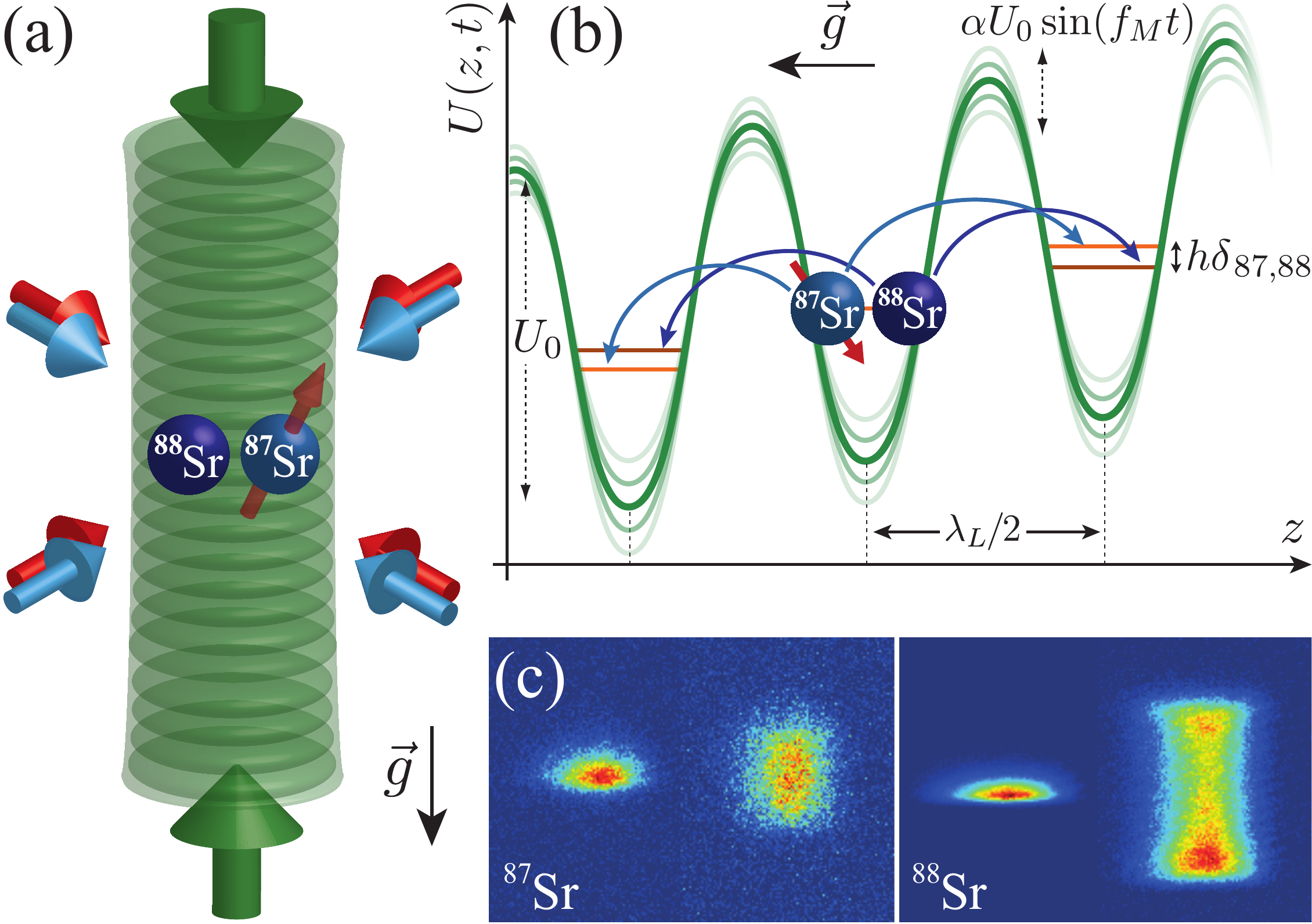}
\caption{(color online) Experimental configuration to test the equivalence principle with Sr atoms. (a) The two isotopes are alternately laser cooled and trapped in a vertical optical lattice. (b) Intraband coherent delocalization of atomic wavepackets is induced by means of amplitude modulation of the optical lattice potential: the difference between the resonant modulation frequencies of the two atomic species $\delta=\nu_{B,88}-\nu_{B,87}$ depends only on their mass ratio and the EP violation parameter $\eta$. (c) Absorbtion images of the \Srf~and \Srb~atomic samples with and without resonant modulation.}
\label{fig:setup}
\end{center}
\end{figure}

In this Letter we report on an experimental comparison of the gravitational interaction for a bosonic isotope of strontium (\Srb) which has zero total spin with that of a fermionic isotope (\Srf) which has a half-integer spin. Sr in the ground state has a $^1$S$_0$ electronic configuration and the total spin corresponds to the nuclear spin { $I$ ($I_{87}=\frac{9}{2}$)}. Gravity acceleration was measured by means of a genuine quantum effect, namely, the coherent delocalization of matter waves in an optical lattice.
%
To compare gravity acceleration for the two Sr isotopes, we confined atomic wave packets  in a vertical off-resonant laser standing wave and induced a dynamical delocalization by amplitude modulation (AM) of the lattice potential~\cite{Alberti2010,Poli2011,Tarallo:2012p4250} at a frequency corresponding to a multiple $\ell$ of the Bloch frequency { $\nu_B = F_g\lambda_L/2h$}, where $h$ is the Planck's constant, $\lambda_L$ is the wavelength of the optical lattice laser  (Fig.~\ref{fig:setup}), and { $F_g$ is the gravitational force on the atomic wavepacket. 

In order to account for anomalous acceleration and spin-dependent gravitational mass, the gravitational potential can be expressed as
\beq\label{eq:anF}
V_{g,A}(z) =\left(1+\beta_A+kS_z\right)m_Agz\, ,
\eeq
where $m_A$ is the rest mass of the atom, $\beta_A$ is the anomalous acceleration generated by a non-zero difference between gravitational and inertial mass due to a coupling with a field with nonmetric interaction with gravity~\cite{Kostelecky2011,Hohensee2013},  $k$ is a model-dependent spin-gravity coupling strength, and $S_z$ is the projection of the atomic spin along gravity direction}.  $k$ can be interpreted as the amplitude of a finite-range mass-spin interaction~\cite{Ni2010}, as a quantum-gravity property of the matter wave field~\cite{Lammerzahl98}, or as a gravitational mass tensor with a spin-dependent component in the Standard Model Extension~\cite{Tasson}. 
The Bloch frequency corresponds to the site-to-site energy difference induced by the gravitational force and, according to the EP, the frequency difference $\delta_{87,88}$ for the two isotopes must depend only on the atomic mass ratio $R_{88,87}=m_{88}/m_{87}$ which is known with a relative uncertainty of $1.5\times10^{-10}$~\cite{Rana:2012p4629}.

The experimental setup was based on a ultrahigh vacuum chamber in which the two Sr isotopes were alternately laser cooled and trapped~\cite{Poli2011}. An oven produced a thermal atomic beam which was slowed in a Zeeman slower and trapped in a magneto-optical trap (MOT) operating on the $^1$S$_0$--$^1$P$_1$ resonance transition at 461 nm. The loading time of the MOT was about 3 s and 7 s for \Srb~and \Srf~atoms, respectively. The temperature was further reduced by a second cooling stage in a ``red'' MOT operating on the $^1$S$_0$--$^3$P$_1$ intercombination transition at 689 nm. In the case of \Srf~atoms, the cooling radiation (cycling on the $F=9/2\rightarrow F'=11/2$ transition) was combined to a second ``stirring'' laser radiation (tuned on the $F=9/2\rightarrow F'=9/2$ transition) to randomize the population of Zeeman sublevels to increase the trapping efficiency~\cite{Mukaiyama03}. The red MOT confined about $5\times 10^6$ \Srb~atoms and $5\times 10^5$ \Srf~atoms with  te
 mperatures of 1 $\mu$K and 1.4 $\mu$K, respectively. The atoms were adiabatically loaded in a vertical optical lattice in 300~$\mu$s. For \Srf, this produced an  unpolarized sample. The lattice potential was generated by a single--mode frequency--doubled $\textrm{Nd:YVO}_4$ laser ($\lambda_L = 532$~nm) delivering up to 1.6~W on the atoms with two counter-propagating beams with a beam waist of about $300~\mu$m. During the gravity measurements the lattice laser frequency was locked to a hyperfine component of molecular iodine by feedback to a piezo-mounted cavity mirror. The single-mode operation of the  laser was monitored using a Fabry-Perot cavity; a self--referenced Ti:Sa optical frequency comb enabled precise calibration of the laser frequency.
The lattice depth $U_0$ was controlled and modulated by two acousto--optical modulators. The atomic cloud was imaged \emph{in situ} at the end of each experiment cycle using resonant absorption imaging on a CCD camera with a spatial resolution of 5~$\mu$m. 

\begin{figure}[tb]
\begin{center}
\includegraphics[width=8. cm]{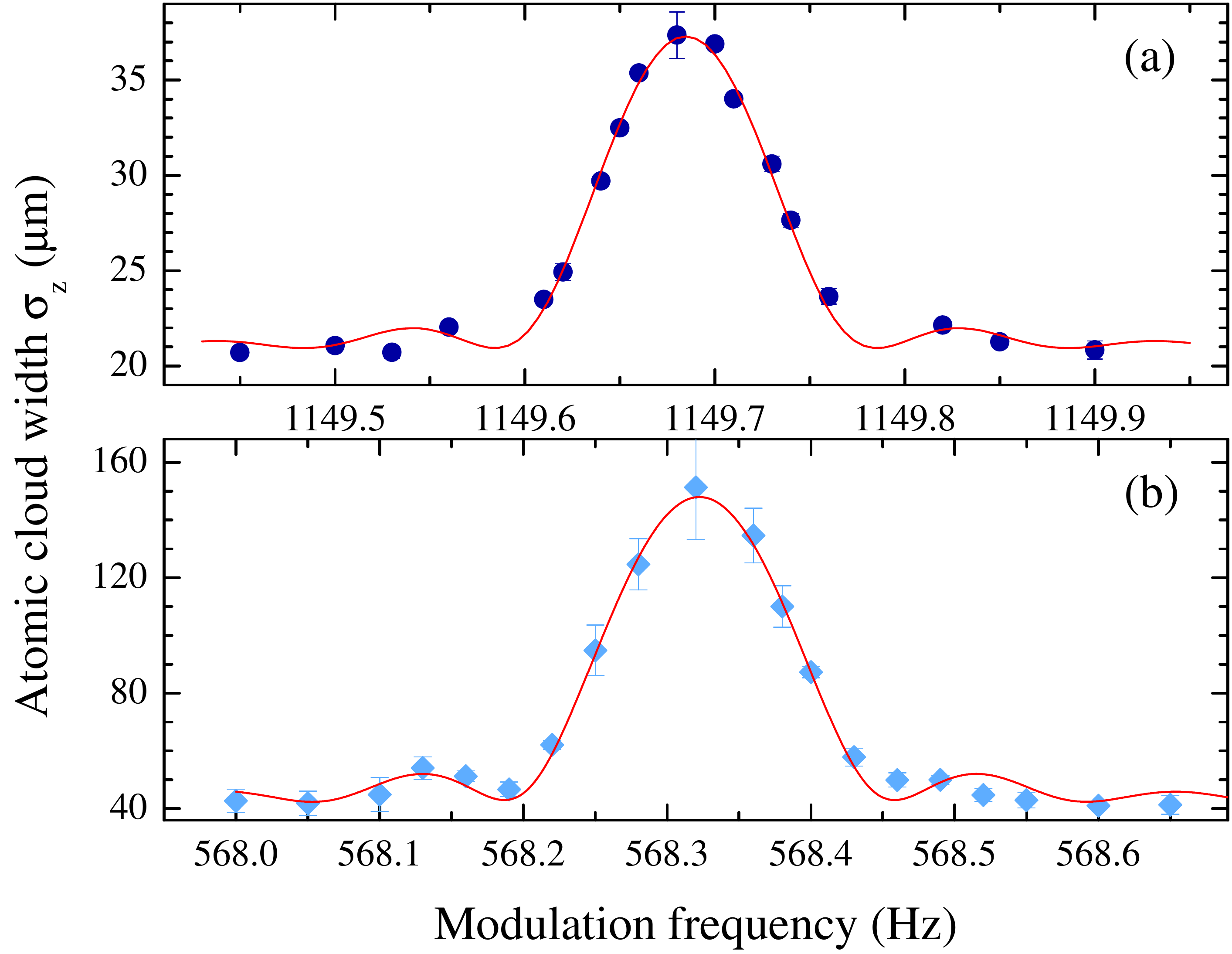}
\caption{
(color online) Typical amplitude modulation spectra and the corresponding least squares best fit function (solid line) for (a) \Srb~($t_{M}$ = 12 s, $\ell$ = 2) and (b) \Srf~($t_{M}$ = 8 s, $\ell$ = 1, $\langle m_F\rangle=0$) atoms. Both the lattice frequency and the lattice beam intensities were kept constant for each pair of measurements, while the modulation depth $\alpha$ was tuned to maximize the signal-to-noise ratio of each spectrum.}
\label{fig:Blochres}
\end{center}
\end{figure}

We measured the Bloch frequency of \Srb~and \Srf~by applying an AM burst to the lattice depth for $t_M=$ 12 s and 8 s at the $\ell=2$ and $\ell=1$ harmonic of $\nu_B$, respectively, and thereafter detecting the resonant broadening of the atomic cloud width $\sigma_z$. A first set of measurements consisted on sweeping the AM frequency $f_M$ to record a full resonance spectrum. The recording time for a whole resonance spectrum was about 1~hour and led to a maximum resolution of $5\times 10^{-7}$ for $\nu_{B,88}$ and $1.6\times10^{-6}$ for $\nu_{B,87}$. A typical resonant tunneling spectrum with the corresponding best fit is shown in Fig.~\ref{fig:Blochres}. The error on the Bloch frequency determination was calculated as the standard error of the fit for each resonance profile.

In this work, we also demonstrated a new method to improve the precision of the measurement of $\nu_B$ and consequently of gravity acceleration by locking the AM oscillator frequency $f_M$ to the Bloch frequency. In analogy to what is done in atomic clocks, $f_M$ can be kept at the top of the resonance spectrum in Fig.~\ref{fig:Blochres} by means of two consecutive AM interrogation cycles at each side of the spectrum. Subsequent demodulation was achieved by computing the difference of the two consecutive measurements of $\sigma_z$, which yielded an odd-symmetry error signal suitable for locking. The slope of the error signal across the resonance was about 0.6 mHz/$\mu$m for typical experimental parameters ($\ell=2$, $t_M=10$ s and {the modulation depth} $\alpha = 6$\% for \Srb, $\ell=1$, $t_M=6.8$ s and $\alpha = 4$\% for \Srf). The Bloch frequency was determined by recording the $f_M$ time series for about 700 s and taking the mean value of the time series.  
\begin{figure}[tb]
\begin{center}
\includegraphics[width=8. cm]{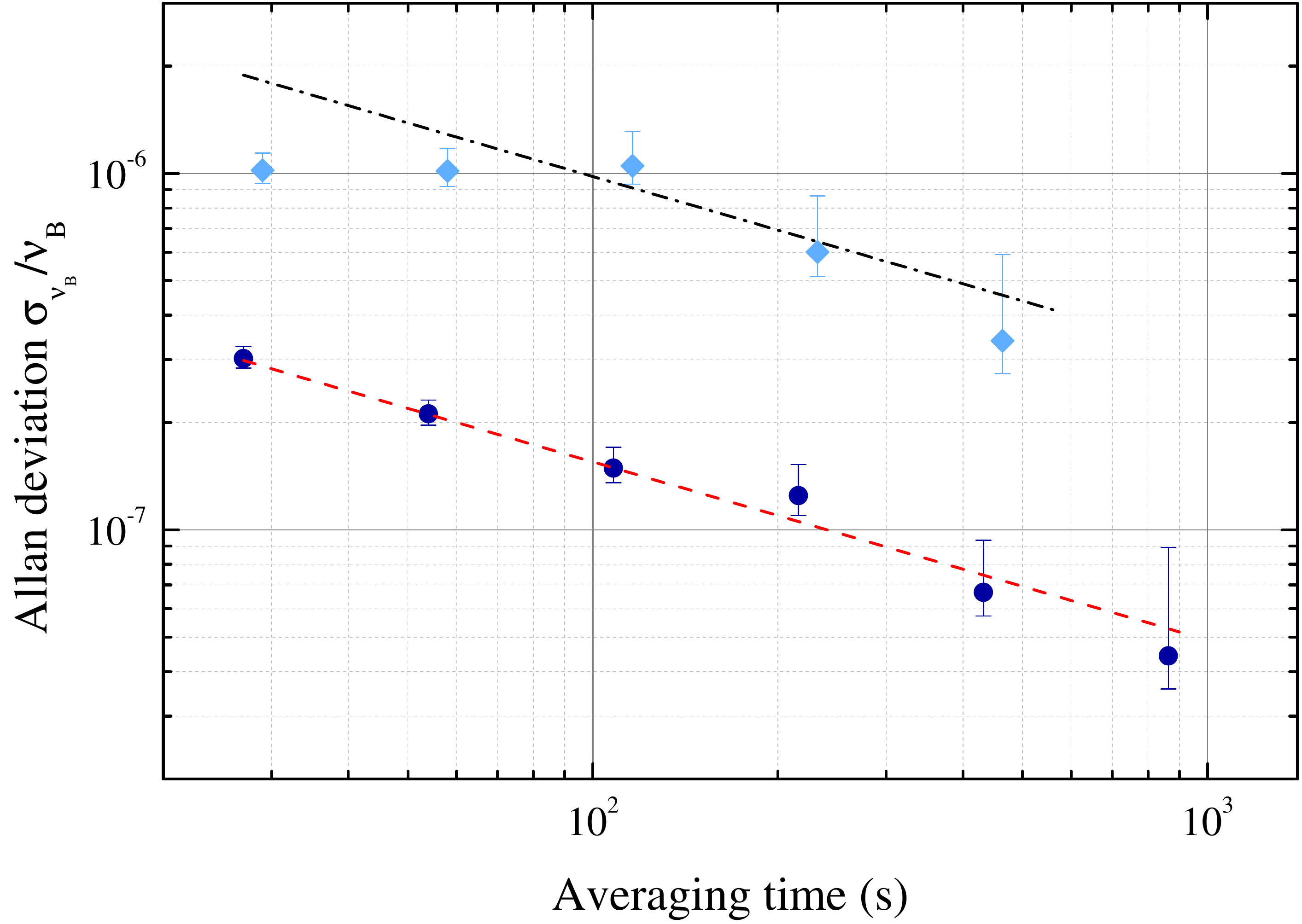}\\
\caption{(color online) Allan deviations of the Bloch frequency measurements for \Srb~(circles) and \Srf~(diamonds) and their corresponding t$^{-1/2}$ asymptotic behaviour (lines) obtained by frequency locking the AM frequency generator to the coherent delocalization resonance, as described in the text.}
\label{fig:AMlock}
\end{center}
\end{figure}
The sensitivity of the Bloch frequency measurement with the new method was characterized by its Allan deviation. Figure~\ref{fig:AMlock} shows the Allan deviation of a set of 101 recorded values of $f_M$ for \Srb~and a set of 42 values for \Srf. In both cases the Allan deviation scales as $t^{-1/2}$ (where $t$ is the measurement time) with sensitivities {at 1~s} of $\sigma_{\nu_{B,88}}=1.5\times10^{-6}\,\nu_{B,88}$ and { $\sigma_{\nu_{B,87}}=9.8\times10^{-6}\,\nu_{B,87}$} respectively. 
This new method, allowed us to improve by more than one order of magnitude the sensitivity in the determination of the frequency of Bloch oscillations (and for gravity acceleration) for \Srb~with respect to our previous results~\cite{Poli2011}, achieving a precision of $5\times 10^{-8}$, while for \Srf~we obtained a precision of $4\times 10^{-7}$. The difference in precision between the two isotopes for both the measurement techniques is due to the reduced signal-to-noise ratio in the absorption profile for \Srf. It is caused by the smaller natural abundance of this isotope and the presence of the 10-level hyperfine manifold that yields a higher Doppler temperature and a smaller (about a factor of 2) absorption cross section due to optical pumping in the imaging process, and, for the frequency lock technique, a slightly higher cycle time (29 s vs 27 s).

\begin{figure}[tb]
\begin{center}
\includegraphics[width=8. cm]{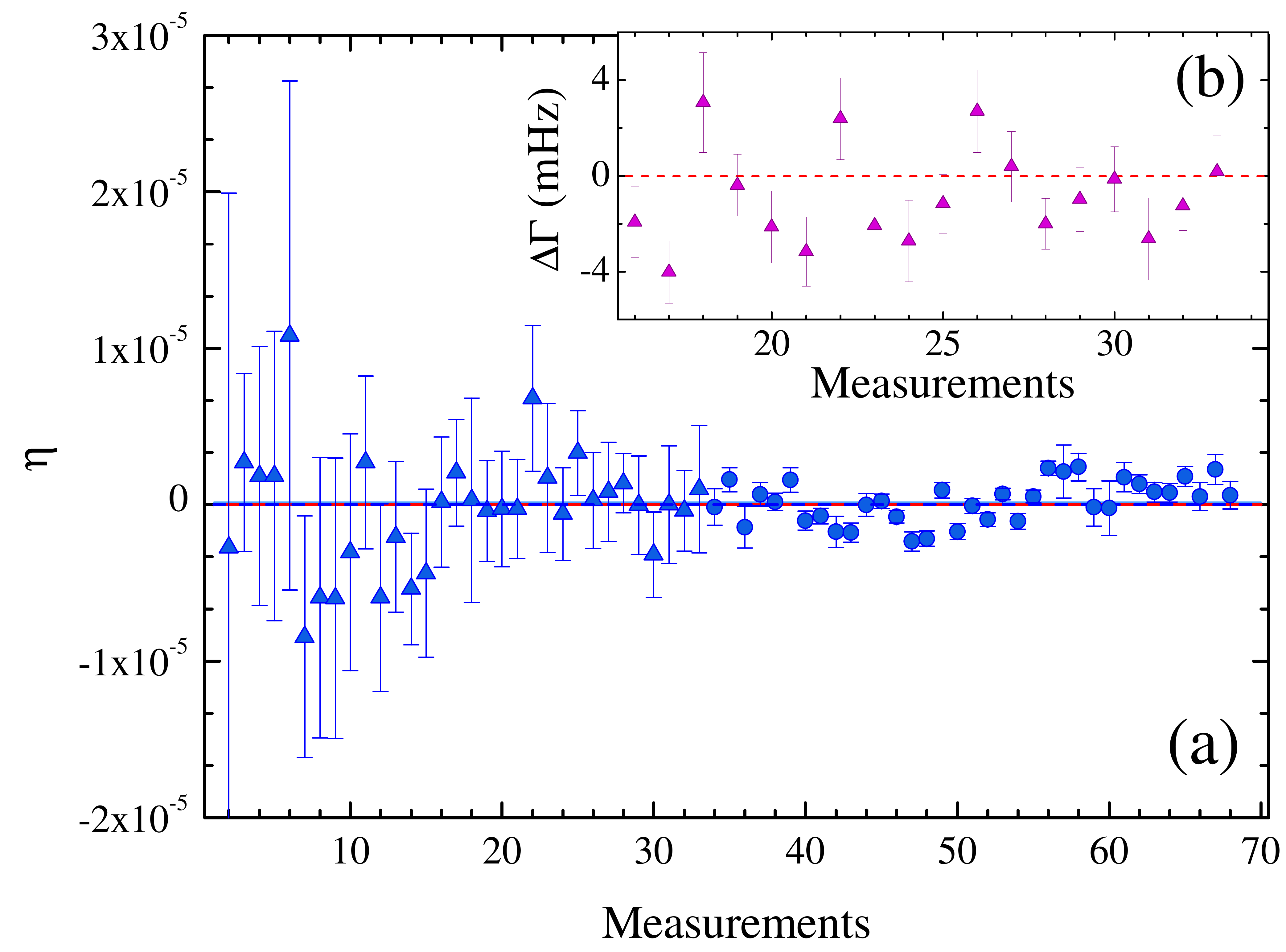}\\
\caption{(color online) Summary of the measurements {for \Srf~and \Srb~Bloch frequency. (a) Measurements} of the $\eta$ parameter by AM resonant tunneling spectra (triangles) and by AM frequency lock (circles). The final weighted mean (blue dashed line) is compared with the null value expected from EP (red line). {(b) Measurements of the resonance linewidth broadening $\Delta\Gamma$ for \Srf~atoms. The dashed (red) line is $\Delta\Gamma=0$} } 
\label{fig:eta}
\end{center}
\end{figure}

Each pair of Bloch frequency measurements was used to determine the
E\"otv\"os ratio~\cite{Eotvos} given by

\beq\label{eq:etaSr} 
\eta \equiv 2\frac{a_{88}-a_{87}}{a_{88}+a_{87}}  =2\frac{\nu_{B,88}-R_{88,87}\ \nu_{B,87}}{\nu_{B,88}+R_{88,87}\ \nu_{B,87}}\, ,
\eeq
where { $a_{i}=2h\nu_{B,i}/m_{i}\lambda_L$ ($i={87,88}$)} are the measured vertical accelerations for the two isotopes.
The data were recorded in $N$ = 68 measurement sessions. Figure~\ref{fig:eta}{ (a)} shows the experimental results for $\eta$, their average value and the comparison with the null value predicted by general relativity. Each point $\eta_i$ is determined with its own error $\sigma_i$ given by the quadratic sum of the statistical error and the uncertainty on the systematic effects. 

In our differential measurement, many systematic errors such as misalignment of the lattice beams, lattice frequency calibration, gravity gradients, and Gouy phase shift, largely cancel and can be neglected at the present level of accuracy. The main contribution to the systematic error in local gravity measurement with trapped neutral atoms arises from the space-dependent lattice depth $U_0(z)$ due to the local intensity gradient of the two interfering Gaussian beams~\cite{Tarallo:2012p4250}. Since we are interested only in the effect of the gravity acceleration upon $\nu_B$, the differential acceleration due to the residual intensity gradient must be removed from the ratio given in (\ref{eq:etaSr}). The correction has been calculated to be

\beq\label{eq:StarkEta}
\Delta\eta_U =\frac{R_{88,87}-1}{(\nu_{B,88}+R_{88,87}\nu_{B,87})/2}\,\frac{\partial_zU_{0}}{2\hbar k_0}\ ,
\eeq
where $k_0=2\pi/\lambda_L$ is the lattice laser wavenumber and we assumed that the difference in the trapping potential due to the dynamic polarizability of the two isotopes is negligible~\cite{middelman}, so that $\partial_zU_0=\partial_zU_{0,88}=\partial_zU_{0,87}$. The expression of the correction in~(\ref{eq:StarkEta}) is then given by the product of the shift of $\nu_{B,88}$ induced by the lattice beam gradient $\Delta\nu_U=\partial_zU_{0}/2\hbar k_0$ and a weight factor $R_{88,87}-1\sim 10^{-2}$ divided by the mean Bloch frequency $(\nu_{B,88}+R_{88,87}\nu_{B,87})/2$. The physical interpretation of (\ref{eq:StarkEta}) is that the acceleration due to the two-photon scattering process producing the confinement in the optical lattice 
has a reduced effect on the differential measurement but does not cancel out. This technical effect affects any EP tests employing an optical {lattice}~\cite{kovac2010}. A precise calibration of the acceleration due to {the} intensity gradient was done by measuring $\nu_{B,88}$ by means of the frequency lock technique. Repeated measurements of $\nu_{B,88}$ were performed with stabilized lattice frequency as {a} function of the total lattice power $P=P_1+P_2+2\varepsilon\sqrt{P_1\ P_2}$, where $P_1$ and $P_2$ is the power sent to the atoms per beam, and $\varepsilon$ is a geometrical correction factor due to the mismatch of the  width of the two beams of order unity. The resulting Bloch frequency shift was $\Delta\nu_U=(\partial\nu_B/\partial P) P = (6.16\pm 0.56)\ 10^{-6}\ P$ Hz/mW corresponding to $\Delta\eta_U \sim 3.6\ 10^{-7}$ for typical operating conditions. The effect of magnetic field gradients in the differential $\nu_B$ measurement was carefully studied. Residual ma
 gnetic field gradients $b=\partial B/\partial z$ were estimated by a precise calibration of the \Srb~red MOT vertical position dynamics to be less than 140 $\mu$T$\cdot$m$^{-1}$. While \Srb~is insensitive to magnetic field gradients at this level of precision~\cite{Poli2011}, the sensitivity of the \Srf~atomic sample depends on the average spin projection $\langle m_F \rangle$. It was estimated by applying a magnetic field gradient up to 210 mT$\cdot$m$^{-1}$ and measuring $\nu_{B,87}$ which resulted in a sensitivity factor $\partial\nu_{B,87}/\partial b = (2\pm 15)$ mHz/(T$\cdot$m$^{-1}$), consistent with a null effect. The effect of tides was estimated {to be} about $1\div0.9\times 10^{-8}$ for a typical time interval of 1 hour between the two $\nu_B$ measurements for the two isotopes. The total systematic uncertainty is thus dominated by the intensity gradient uncertainty at the level of $3\times10^{-8}$, while a residual lattice frequency error due to the frequency lock 
 precision has been estimatedÊ{ to be} lower than $1\times10^{-8}$.

The final result for the $\eta$ parameter is $\eta =(0.2\pm 1.6)\times10^{-7}$, where the uncertainty corresponds to the standard deviation of the weighted mean { $\sigma_{\bar{\eta}}=\sqrt{1/\sum_N(\sigma_i^{-2})}$, corrected by the reduced chi-square ($\chi^2/(N-1)=2.78$)}. {In the case of unpolarized \Srf~atoms, the mean contribution of $kS_z$ is zero and $\eta = \beta_{88}-\beta_{87}$}. This result can be interpreted in terms of the EP violation parameters for the fundamental constituents of the two atoms, according to different parametrizations~\cite{Damour96,Hohensee2013}{, and it sets a $10^{-7}$ direct bound on the boson-to-fermion gravitational constant ratio $f_{BF}$ from being different from 1~\cite{Barrow}}. {On the other hand, each \Srf~spin component $S_z=I_{z}$ will feel different gravitational forces due to different spin-gravity coupling, as in the case of a magnetic field gradient, resulting in a broadening of the frequency response shown in Fig.~\ref{fig:Bl
 ochres}. We analyzed a set of \Srf~AM resonant tunneling spectra used for the determination of $\eta$. The residual deviations of the measured linewidth $\Gamma$ from the Fourier linewidth, after removing systematic broadening mechanisms such as the ones due to the two-body collisions and the residual magnetic field gradients, are shown in Fig.\ref{fig:eta}(b). The measured residual broadening $\Delta\Gamma = 0.4\pm0.5$(stat.)$\pm0.8$(sys.) mHz sets an upper limit on the spin-gravity coupling $\Delta\Gamma = 2I_{87}k\ell\nu_{B,87}$, resulting on a spin-gravity coupling strength
\[
k = (0.5 \pm 1.1)\,10^{-7} \,.
\]
Since the nucleus of  \Srf~has 9 valence neutrons,
this result  also sets a limit of $10^{-7}$ for anomalous acceleration and spin-gravity coupling for the neutron either as a difference in the gravitational mass depending on the spin direction, which was previously limited at $10^{-23}$~\cite{Venema1992}, or as a coupling to a finite-range Leitner-Okubo-Hari Dass interaction, which was limited to less than 10 at 30 mm~\cite{Ni2010}.}

In conclusion, we performed a {quantum} test of EP for the bosonic \Srb~isotope which has no spin vs the fermionic \Srf~isotope which has a half-integer spin by coherent control of the atomic motion in an optical lattice under the effect of gravity. 
{We obtained upper limits of $\sim 10^{-7}$ for pure inertial effects and for a possible spin-gravity coupling}. The present results can set bounds for previously unmeasured parameters of the Standard Model Extension~\cite{Kostelecky2011,Tasson}.
{Further} enhancements in  sensitivity will require the development of higher transferred momentum atom interferometry schemes for Sr atoms, and simultaneous probing of the two isotopes \cite{Poli2005}. {Short-distance measurements ($r\leq 1$ cm) with $10^{-8}\nu_B$ precision can lower the limit of monopole-dipole interaction constants $g_pg_s$ of nine orders of magnitude~\cite{Kimball2013}}. At the same time, Sr optical clocks are showing impressive advances in stability and accuracy with the possibility of building compact and transportable systems \cite{Poli2013}. Results from a network of Sr optical clocks already set a limit to the coupling of fundamental constants to gravity~\cite{Blatt08}. It is possible then to envisage a future experiment in space where a Sr interferometer and a Sr optical clock would be operated at their limit performances to realize stringent tests of general relativity~\cite{Tino14}.
%



\begin{acknowledgments}
This project has received funding from INFN (MAGIA experiment) and from the EUÕs 7th FP under grant agreement n. 250072. We thank S. Capozziello, J. Tasson and T. Zelevinsky for a critical reading of the manuscript and for useful suggestions.
\end{acknowledgments}


\end{document}